\documentclass[%
twocolumn,
amsmath,amssymb,
aps,
pre,
longbibliography
]{revtex4-1}

\usepackage{graphicx,epsfig,epstopdf}
\usepackage{amsmath}
\usepackage{color}
\usepackage{caption}
\usepackage{subfigure}
\usepackage{array}
\usepackage{tabularx}
\usepackage{multirow}
\usepackage{gensymb}
\usepackage{soul}

\usepackage{bm,hyperref}
\hypersetup{pdfpagemode=UseNone}
\usepackage{lipsum}
\usepackage{amsfonts}
\usepackage{amssymb}
\usepackage{dcolumn}
\usepackage{textcomp}
\newcolumntype{L}[1]{>{\raggedright\arraybackslash}p{#1}}
\newcolumntype{C}[1]{>{\centering\arraybackslash}p{#1}}
\newcolumntype{M}[1]{>{\centering\arraybackslash}m{#1}}

\def\e{\begin{equation}}
\def\f{\end{equation}}
\def\_#1{{\bf #1}}

\def\.{\cdot}

\def\=#1{\overline{\overline #1}}

\def\-#1{{\bf #1}}

\begin{document}

\title{A novel analytical method for analysis of electromagnetic scattering from inhomogeneous spherical structures using duality principles}

\author{
M. Kiani$^{\ast}$, A. Abdolali, and  M. Safari
}
 
\affiliation{Department of Electrical Engineering, Iran University of Science and Technology, Narmak, Tehran, Iran\\
~~$^\ast${\rm corresponding author}
}

\begin{abstract}
In this article, a novel analytical approach is presented for the analysis of electromagnetic (EM) scattering from radially inhomogeneous spherical structures (RISSs) based on the duality principle. According to the spherical symmetry, similar angular dependencies in all the regions are considered using spherical harmonics. To extract the radial dependency, the system of differential equations of wave propagation toward the inhomogeneity direction is equated with the dual planar ones. A general duality between electromagnetic fields and parameters and scattering parameters of the two structures is introduced. The validity of the proposed approach is verified through a comprehensive example. The presented approach substitutes a complicated problem in spherical coordinate to an easy, well posed, and a previously solved problem in planar geometry. This approach is valid for all continuously varying inhomogeneity profiles. One of the major advantages of the proposed method is the capability of studying two general and applicable types of RISSs. As an interesting application, a new class of lens antenna based on the physical concept of the gradient refractive index material is introduced. The approach is used to analyze the EM scattering from the structure and validate strong performance of the lens.
\end{abstract}

\maketitle

\section{Introduction}

Analysis of electromagnetic (EM) scattering from spherical structures containing complex media, such as metamaterials [1,2], anisotropic [3-8], bi-anisotropic [9-11], dispersive [12], and inhomogeneous media [13-22] has always been among the most interesting topics for researchers because of their complicated constitutive relations, and therefore having more involved parameters and degrees of freedom in design.
Meanwhile, inhomogeneous media are of the great importance owing to their extensive use in lens antenna design [23-26] and transparency [5,27].
 In addition, according to the presented applications of inhomogeneous planar and cylindrical structures in [28-32], using these media in spherical structures could also result in potentially interesting applications. Along with the abovementioned investigations, several recent researches in the field of inhomogeneous spherical structures show the significance of analyzing EM scattering from these structures [18-22]. Most studies in the field of spherical structures investigated the case of radial inhomogeneity [14-22] or anisotropy [33]. Although [13] is one of the main investigations for analysis of EM scattering from 3D inhomogeneous sphere, which is presented for the general case. In all the studies in the field of radially inhomogeneous media, to make use of the advantage of spherical symmetry, similar known angular dependency has been used based on the spherical harmonics. These harmonics are orthogonal over a spherical surface. The reason for the several investigations in this field is to present different approaches to extract the radial dependencies of EM fields in the inhomogeneous media.  Preliminary investigations on analysis of EM scattering from radially inhomogeneous spherical structures (RISSs) used approximate methods, namely Born approximation and multilayer equivalent [14-17]. Contrary to these investigations, which are based on approximations, recent studies based on hybrid (analytical-numerical) methods for analysis of RISSs were presented in the literature. In [18, 19], orthogonal Dini-type basis functions are used to solve the electric field volume integral equation of an inhomogeneous gyroelectric sphere.
In [20], using the spherical symmetry, the 3D integral equation for inhomogeneous dielectric sphere was converted to a 1D integral equation in radial direction, which simplified the analysis process and reduced calculation weight. 
In [21, 22], an approximate method is proposed for the analysis of EM scattering from an inhomogeneous sphere with a size that is smaller than the vacuum wavelength, which allows using only the two lowest spherical harmonics.
In this study, a novel analytical approach for analysis of EM scattering from RISSs using duality principle is proposed. Similar angular dependencies in the researches mentioned are used, which are based on spherical harmonics. 
The system of differential equations describing the EM propagation is derived toward the inhomogeneity direction accurately and directly from the Maxwell equations. To solve the system of equations, the results of the dual planar problem, which is far simpler and previously solved, are used.  The proposed duality is based on equating the wave differential equations in the direction of inhomogeneity and boundary conditions of the mentioned two structures, which are the key elements for solving each electromagnetic problem. The validity of the proposed duality is verified for a special case of RISS, which has an analytical exact solution.
The EM problems in spherical structures are known to be difficult. Using inhomogeneous medium in such structures can significantly increase the complexity of the problem. Moreover, providing a mathematical method for solving the governing equations require intensive calculations and substantial processing for these problems.
Combining the mentioned difficulties turns the analysis of RISSs into a very complicated problem with extensive analysis process. The proposed duality overcomes the mentioned complexities and difficulties by introducing a dual planar structure for any IRSSs, which is far simpler, well posed, and previously solved. This study results in an accurate solution, using a simple, new, and analytical approach.
RISSs are divided into two general (applicable) categories, namely an arbitrary spherical core coated with an inhomogeneous layer and an inhomogeneous sphere. Contrary to the mentioned studies in literature, one of the advantages of the current investigation is that both applicable categories of RISSs are analyzed.
In the last section, an inhomogeneous dielectric sphere is introduced as a new class of lens antenna based on the basic concept of gradient refractive index material. Comparing the results of the proposed method and other conventional approaches confirm the applicability and strong performance of proposed lens antenna relative to the Luneburg lens. The proposed approach can be used for other interesting applications such as designing radar absorbers, cloaks, and radomes in future researches.

\section{PROBLEM DEFINITION}
\begin{figure}[!t]
\centering
\includegraphics[width=2.5in]{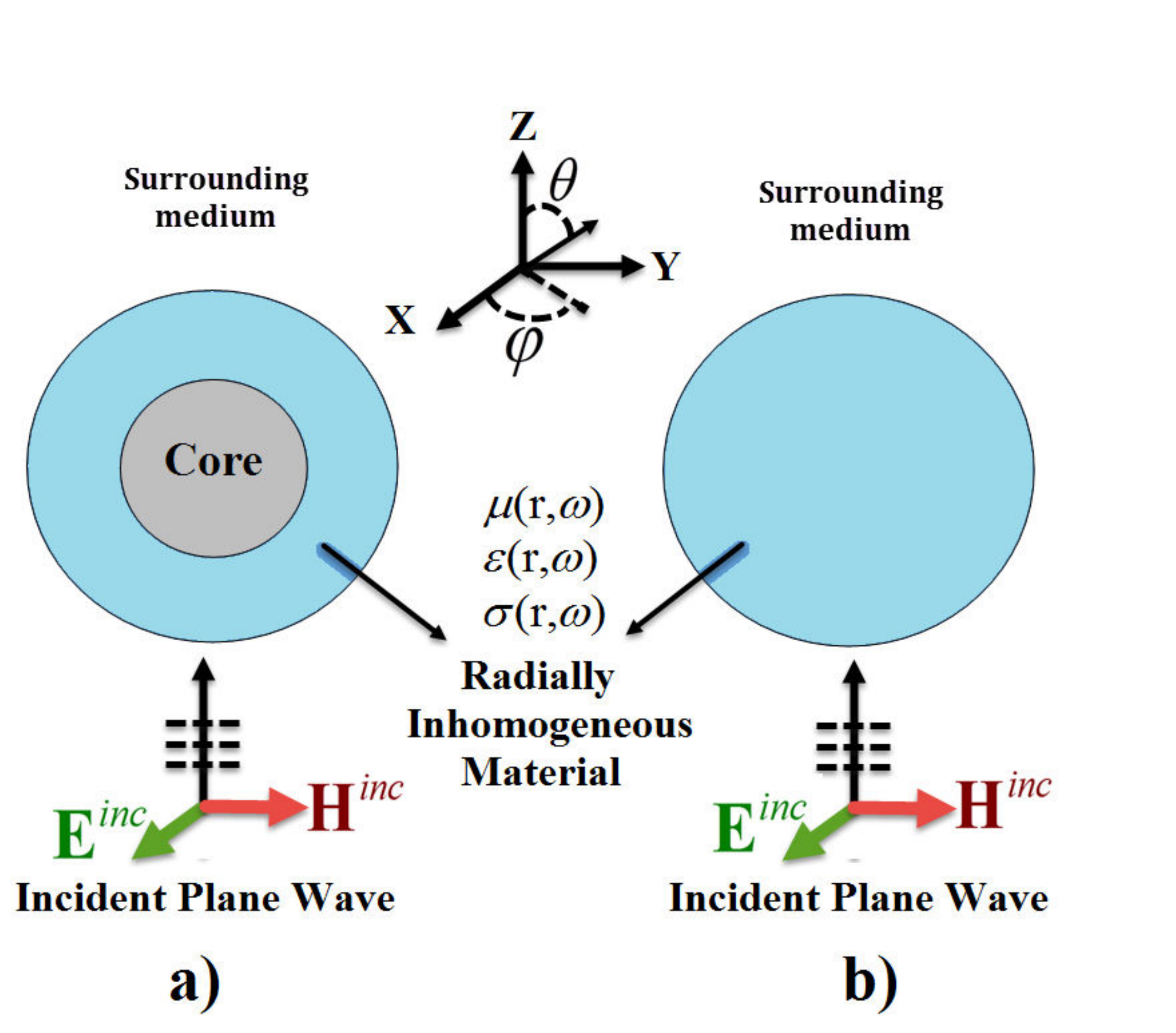}
\DeclareGraphicsExtensions.
\caption{RISS: sphere core coated with a radially inhomogeneous layer (a) radially inhomogeneous sphere (b)}
\label{fig_1}
\end{figure}
Consider a radially inhomogeneous spherical structure as illustrated in Fig. 1. The mentioned structure is divided into two general and practical categories, namely, a spherical core coated with an inhomogeneous layer and an inhomogeneous sphere, which are introduced in Fig. 1 a and b, respectively. The core of the structure in the first category can be a Perfect Electrical Conductor (PEC), a Perfect Magnetic Conductor  (PMC), impedance surface, dielectric, or metamaterial (see Fig. 1 a). It is assumed that the plane wave with ${{E}_{x}}$ component and amplitude of the electric field ${{E}_{0}}$ propagating toward the positive z direction illuminates the mentioned structures. The transverse components of incident fields can be represented by an infinite sum of spherical wave functions in the following manner [34]:
\begin{equation}
\begin{split}
E_{\theta }^{inc}=
-{E}_{0}\frac{\cos\phi}{{k}_{0}r}
\sum\nolimits_{n=1}^{\infty}{j}^{-n}\frac{2n+1}{n(n+1)}\left({\begin{array}{cc} j\hat{J_{n}^{'}}({{k}_{0}}r)\frac{dP_{n}^{1}(\cos \theta )}{d\theta}  \\ +\hat{{J}_{n}}({{k}_{0}}r)\frac{P_{n}^{1}(\cos \theta )}{\sin \theta } \end{array}}\right)
\end{split}
\end{equation}
\begin{equation}
\begin{split}
E_{\phi }^{inc}=
{E}_{0}\frac{\sin\phi}{{k}_{0}r}
\sum\nolimits_{n=1}^{\infty}{j}^{-n}\frac{2n+1}{n(n+1)}\left({\begin{array}{cc} j\hat{J_{n}^{'}}({{k}_{0}}r)\frac{P_{n}^{1}(\cos\theta)}{\sin\theta}  \\ +\hat{{J}_{n}}({{k}_{0}}r)\frac{dP_{n}^{1}(\cos \theta )}{d\theta } \end{array}}\right)
\end{split}
\end{equation}
\begin{equation}
\begin{split}
H_{\theta }^{inc}=
-\frac{{E}_{0}}{\eta_{0}}\frac{\sin\phi}{{k}_{0}r}
\sum\nolimits_{n=1}^{\infty}{j}^{-n}\frac{2n+1}{n(n+1)}\left({\begin{array}{cc} j\hat{J_{n}^{'}}({{k}_{0}}r)\frac{dP_{n}^{1}(\cos \theta )}{d\theta}  \\ +\hat{{J}_{n}}({{k}_{0}}r)\frac{P_{n}^{1}(\cos \theta )}{\sin \theta } \end{array}}\right)
\end{split}
\end{equation}
\begin{equation}
\begin{split}
H_{\phi}^{inc}=
-\frac{{E}_{0}}{\eta_{0}}\frac{\cos\phi}{{k}_{0}r}
\sum\nolimits_{n=1}^{\infty}{j}^{-n}\frac{2n+1}{n(n+1)}\left({\begin{array}{cc} j\hat{J_{n}^{'}}({{k}_{0}}r)\frac{P_{n}^{1}(\cos\theta)}{\sin\theta}  \\ +\hat{{J}_{n}}({{k}_{0}}r)\frac{dP_{n}^{1}(\cos \theta )}{d\theta } \end{array}}\right)
\end{split}
\end{equation} 
Where ${{k}_{0}}$and ${{\eta }_{0}}$ are the wave number and intrinsic impedance of the surrounding media, respectively. According to the physical concept of the problem, as a portion of the incident EM wave power penetrates into the inhomogeneous layer, the problem of EM wave propagation in the inhomogeneous media must be addressed.
Because the inhomogeneity is only toward the radial direction, the problem has spherical symmetry. Therefore, similar to previous studies in the field of radially inhomogeneous media [18-22], the angular dependency can be shown with respect to spherical harmonics in the following manner. These harmonics are orthogonal over a spherical surface.
\begin{equation}
{{\psi }_{m,n}}(\theta ,\phi )=P_{n}^{m}(\cos \theta ){{e}^{jm\phi }}
\end{equation}
According to the definition of incident plane wave with ${{E}_{x}}$ component (m=1 is assumed) as the inhomogeneous medium that is assumed to be linear, the general form of spatial dependency of the auxiliary vector potentials for TEr and TMr modes is shown as follows, based on spherical harmonics.
\begin{equation}
F_{r}^{inh}(r,\theta ,\phi ,\omega )=\sin \phi \sum\nolimits_{n=1}^{\infty }{F_{r}^{inh,n}(r,\omega )P_{n}^{1}(\cos \theta )}
\end{equation}
\begin{equation}
A_{r}^{inh}(r,\theta ,\phi ,\omega )=\cos \phi \sum\nolimits_{n=1}^{\infty }{A_{r}^{inh,n}(r,\omega )P_{n}^{1}(\cos \theta )}
\end{equation}
Where $A_{r}^{inh,n}(r,\omega )$ and $F_{r}^{inh,n}(r,\omega )$ represent the radial and frequency dependency of auxiliary potentials in a radially inhomogeneous medium and satisfy the following equations (see Appendix A).
\begin{equation}
\begin{split}
 &\mu(r,\omega ){{\varepsilon }_{comp}}(r,\omega )\frac{\partial }{\partial r}\left( \frac{1}{\mu (r,\omega )}\frac{\partial }{\partial r}(\frac{F_{r}^{inh,n}(r,\omega )}{{{\varepsilon }
_{comp}}(r,\omega )}) \right) \\
  &+{{\omega }^{2}}\mu (r,\omega ){{\varepsilon }_{comp}}(r,\omega )F_{r}^{inh,n}(r,\omega ) \\
&-\frac{n(n+1)}{{{r}^{2}}}F_{r}^{inh,n}(r,\omega )=0
\end{split}
\end{equation}
\begin{equation}
\begin{split}
{\varepsilon}_{comp}(r,\omega )\mu (r,\omega )\frac{\partial}{\partial r}\left( \frac{1}{{{\varepsilon }_{comp}}(r,\omega )}\frac{\partial }{\partial r}(\frac{A_{r}^{inh,n}(r,\omega )}{\mu (r,\omega )}) \right) \\ 
 +{{\omega }^{2}}\mu (r,\omega ){{\varepsilon }_{comp}}(r,\omega )A_{r}
 -\frac{n(n+1)}{{{r}^{2}}}A_{r}^{inh,n}(r,\omega )=0 
\end{split}
\end{equation}
These equations do not have an analytical solution except for some special cases.
The general form of spatial dependency for the transverse components of EM fields can be defined in the following way:
\begin{align}
E_{r}^{inh}=\cos \phi \sum\nolimits_{n=1}^{\infty }{E_{r}^{inh,n}(r,\omega )P_{1}^{n}(\cos \theta )} \\
E_{\theta }^{inh}=\cos \phi \sum\nolimits_{n=1}^{\infty} 
\left(\begin{array}{cc} E_{\theta ,A}^{inh,n}(r,\omega )\frac{dP_{1}^{n}(\cos \theta )}{d\theta } \\
+E_{\theta ,F}^{inh,n}(r,\omega )\frac{P_{1}^{n}(\cos \theta )}{\sin\theta} \end{array} \right)\\
E_{\phi }^{inh}=\sin \phi \sum\nolimits_{n=1}^{\infty} 
\left(\begin{array}{cc} E_{\phi ,A}^{inh,n}(r,\omega )\frac{P_{1}^{n}(\cos \theta )}{\sin\theta } \\
+E_{\phi ,F}^{inh,n}(r,\omega )\frac{dP_{1}^{n}(d\theta )}{\sin\theta} \end{array} \right)\\
H_{r}^{inh}=\sin \phi \sum\nolimits_{n=1}^{\infty }{H_{r}^{inh,n}(r,\omega )P_{1}^{n}(\cos \theta )}\\
H_{\theta }^{inh}=-\sin \phi \sum\nolimits_{n=1}^{\infty} 
\left(\begin{array}{cc} H_{\theta ,A}^{inh,n}(r,\omega )\frac{P_{1}^{n}(\cos \theta )}{\sin\theta } \\
+H_{\theta ,F}^{inh,n}(r,\omega )\frac{dP_{1}^{n}(d\theta )}{\sin\theta} \end{array} \right)\\
H_{\phi }^{inh}=-\cos \phi \sum\nolimits_{n=1}^{\infty} 
\left(\begin{array}{cc} H_{\phi ,A}^{inh,n}(r,\omega )\frac{dP_{1}^{n}(\cos \theta )}{d\theta } \\
+H_{\phi ,F}^{inh,n}(r,\omega )\frac{P_{1}^{n}(\cos \theta )}{\sin\theta} \end{array} \right)
\end{align}
Where 
\begin{align}
H_{\phi ,A}^{inh,n}(r,\omega)=H_{\theta ,A}^{inh,n}(r,\omega )\\
H_{\phi ,F}^{inh,n}(r,\omega )=H_{\theta ,F}^{inh,n}(r,\omega )\\
E_{\phi ,A}^{inh,n}(r,\omega )=E_{\theta ,A}^{inh,n}(r,\omega )\\ 
E_{\phi ,F}^{inh,n}(r,\omega )=E_{\theta ,F}^{inh,n}(r,\omega)
\end{align}
The extraction of the above equations from auxiliary potentials and the relation between EM fields are described in detail in Appendix A.$E_{\phi ,F}^{inh,n}(r,\omega )$,$E_{\phi ,A}^{inh,n}(r,\omega )$,$H_{\phi ,F}^{inh,n}(r,\omega )$, and $H_{\phi ,A}^{inh,n}(r,\omega )$ represent the radial and frequency dependency of transverse fields. In the following, the extraction of the mentioned dependencies as the key parameters of the problem is discussed. Replacing the electromagnetic fields by Maxwell equations and equating the components with the same angular variation, the differential equations system of wave propagation toward the direction of inhomogeneity is derived.
\begin{equation}
\begin{split}
& \frac{\partial }{\partial r}E_{\phi ,A}^{inh,n}(r,\omega)+\frac{E_{\phi ,A}^{inh,n}(r,\omega )}{r} \\ 
&=\left( -\frac{n(n+1)}{j\omega {{r}^{2}}{{\varepsilon }_{comp}}(r,\omega )}-j\omega \mu (r,\omega ) \right)H_{\phi ,A}^{inh,n}(r,\omega ) 
\end{split}
\end{equation}
\begin{equation}
\begin{split}
 & \frac{\partial}{\partial r}E_{\phi ,F}^{inh,n}(r,\omega )+\frac{E_{\phi ,F}^{inh,n}(r,\omega )}{r} \\ 
&=-j\omega \mu (r,\omega )H_{\phi ,F}^{inh,n}(r,\omega) 
\end{split}
\end{equation}
\begin{equation}
\begin{split}
&\frac{\partial }{\partial r}H_{\phi ,F}^{inh,n}(r,\omega )+\frac{H_{\phi ,F}^{inh,n}(r,\omega )}{r} \\ 
&=( -j\omega {{\varepsilon}_{comp}}(r,\omega )-\frac{n(n+1)}{j\omega {{r}^{2}}\mu (r,\omega )})E_{\phi ,F}^{inh,n}(r,\omega ) 
\end{split}
\end{equation}
\begin{equation}
\begin{split}
&\frac{\partial }{\partial r} H_{\phi ,A}^{inh,n}(r,\omega)+\frac{H_{\phi ,A}^{inh,n}(r,\omega )}{r} \\ 
& =-j\omega {{\varepsilon }_{comp}}(r,\omega )E_{\phi ,A}^{inh,n}(r,\omega )
\end{split}
\end{equation}
The differential equation system in (20)-(23) has no exact solution except for some special cases.
Similar to previous studies in the field of planar and cylindrical structures [28-32], the process of derivation of EM fields’ variation in the direction of inhomogeneity are based on the solution of the wave propagation system of differential equations. Therefore, the difference between equations form and the process of extracting them in comparison with the recent studies in the field of radially inhomogeneous spherical structures is obvious and expected [18-22]. However, it is of significant importance to mention that all the relations and equations in this study are derived directly and exactly from Maxwell equations, which ensures a converged unique solution for the problem. It is clear that the process of solving the system of equations and extracting the radial dependency of the fields is difficult. In addition, proposing a mathematical solution for the mentioned system requires extensive and complex calculations, which increase the difficulty of the problem. To simplify the analysis process, a novel approach based on the duality principle is proposed. This approach uses the results of the dual inhomogeneous planar problem, which is far simpler and more well-posed than the main problem in spherical coordinate. In addition, various analytical methods are presented in the literature to solve the dual problem. In the next section, the equations of the dual planar structure and duality relations are presented.
\subsection{The dual IPL}
\begin{figure}[!t]
\centering
\includegraphics[width=2.5in]{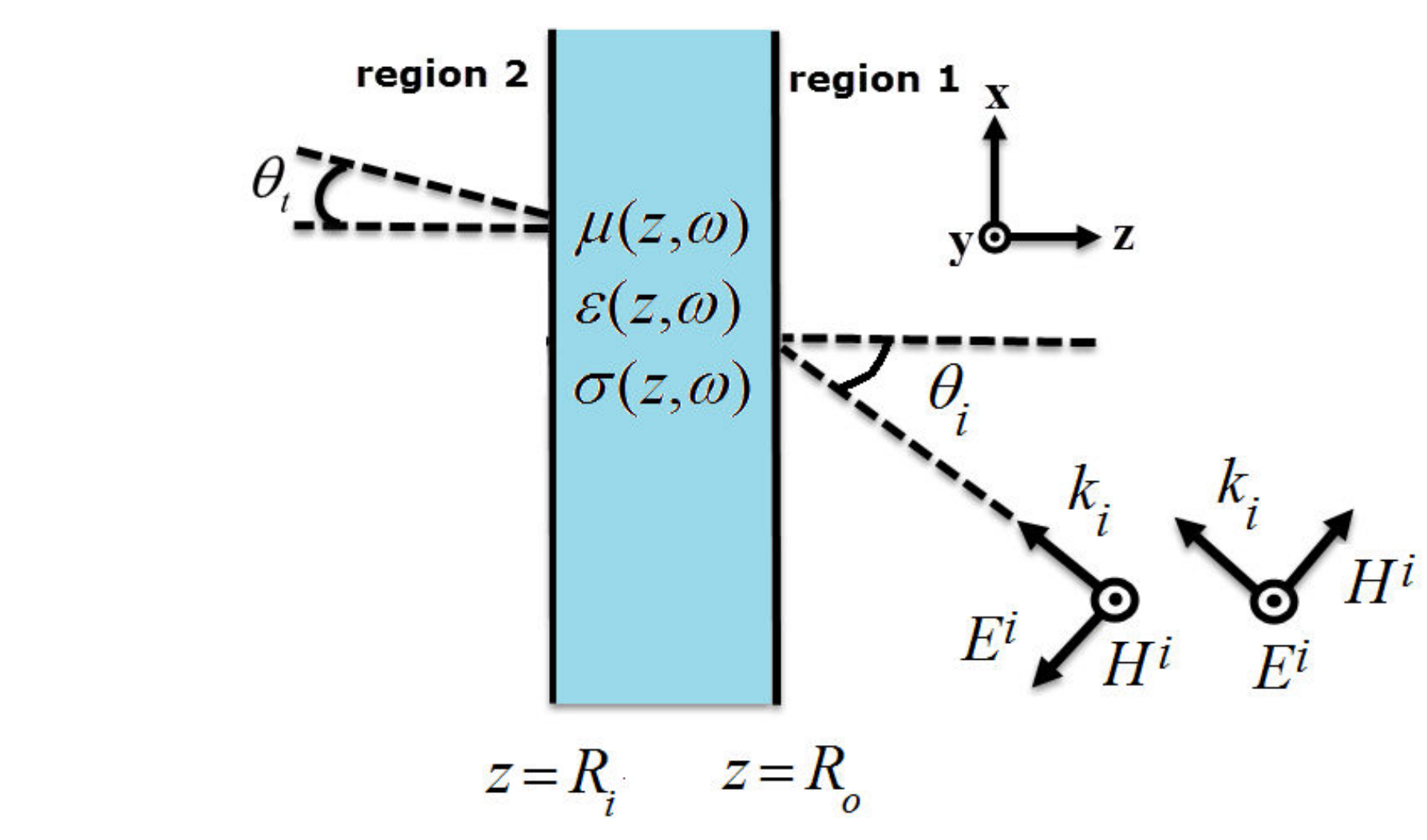}
\DeclareGraphicsExtensions.
\caption{The Dual IPL}
\label{fig_2}
\end{figure}
Consider an inhomogeneous planar layer (IPL) as illustrated in Fig. 2. The plane wave propagates toward the negative z direction and the structure is obliquely illuminated. The system of differential equation describing EM wave propagation in inhomogeneous layer is obtained similar to [35, 36] in the following form: \\
for ${TE}^{z}$:
\begin{equation}
\frac{\partial {{E}_{y}}(z,\omega )}{\partial z}=j\omega \mu (z,\omega ){{H}_{x}}(z,\omega )
\end{equation}
\begin{equation}
\frac{\partial {{H}_{x}}(z,\omega )}{\partial z}=\left( j\omega {{\varepsilon }_{comp}}(z,\omega )+\frac{{{k}_{x}}^{2}}{j\omega \mu (z,\omega )} \right){{E}_{y}}(z,\omega )
\end{equation}
for ${TM}^{z}$:
\begin{equation}
\frac{\partial {{H}_{y}}(z,\omega )}{\partial z}=-j\omega {{\varepsilon }_{comp}}(z,\omega ){{E}_{x}}(z,\omega)
\end{equation}
\begin{equation}
\frac{\partial {{E}_{x}}(z,\omega )}{\partial z}=-\left( \,j\omega \mu (z,\omega )+\frac{{{k}_{x}}^{2}}{j\omega {{\varepsilon }_{comp}}(z,\omega )} \right){{H}_{y}}(z,\omega )
\end{equation}
Where ${{k}_{x}}={{k}_{0}}sin\theta$ and ${{k}_{0}}$ is the wave number of the region 1 (see Fig. 2). To derive the duality, similar to the spherical structure, inhomogeneous planar layer is located at $z=[{{R}_{i}},{{R}_{o}}]$. In addition, the regions 1 and 2 in Fig. 2 have the same EM parameters with the surrounding medium and core in the RISS, respectively. For PEC, PMC, and impedance cores similar boundary conditions are considered at $z={{R}_{i}}$.
By comparing (20)-(23) with (24)-(27), it can be clearly seen that the equations describing the components of EM field related to A and F potentials in RISS and those of planar structure for TEz and TMz polarizations are similarly uncoupled. This fact is the primary basis of the novel duality establishment between inhomogeneous spherical and planar structures is presented in Table 1. This duality is explained as an analytical approach for analysis of EM scattering.

In the analysis of scattering problems, satisfaction of boundary conditions is as important as extraction and solving the wave equations. Therefore, in the following, the duality between boundary conditions for both structures is presented. For this purpose, the general forms of vector potentials in the homogeneous surrounding medium and dielectric, or metamaterial core are presented in the following manner:\\
\begin{align}
A_{r}^{s}=\frac{{{E}_{0}}\sin \phi }{{{\eta }_{0}}\omega }\sum\nolimits_{n=0}^{\infty }{{{a}_{n}}{\hat{H}_{n}^{(2)}}\,({{k}_{0}}r)P_{n}^{1}(\cos \theta )} \\
F_{r}^{s}=\frac{{{E}_{0}}\sin \phi }{\omega }\sum\nolimits_{n=0}^{\infty }{{{b}_{n}}{\hat{H}_{n}^{(2)}}\,({{k}_{0}}r)P_{n}^{1}(\cos \theta )} \\
A_{r}^{t}=\frac{\cos \phi }{\omega }\sum\nolimits_{n=1}^{\infty }{{{c}_{n}}{\hat{{{J}_{n}}}}\,({{k}_{d}}r)P_{n}^{1}(\cos \theta )} \\
F_{r}^{t}=\frac{\sin \phi }{\omega }\sum\nolimits_{n=0}^{\infty }{{{d}_{n}}{\hat{{{J}_{n}}}}\,({{k}_{d}}r)P_{n}^{1}(\cos \theta )} 
\end{align}
Where a ${{k}_{d}}$ is the wave number of dielectric or metamaterial core, s and t superscripts represent scattered fields from the structure and fields that are propagated in the core, respectively.
According to the spherical symmetry of the problem, similar angular variation is considered in the above equations, and the radially inhomogeneous medium, which is based on spherical harmonics, are both similar.
\begin{table}[!h]
\renewcommand{\arraystretch}{1.9}

\caption{Comprehensive duality between electromagnetic fields, electromagnetic parameters}
\label{table_1}
\centering

\begin{tabular}{|c||c|}
\hline
IPL & ISL\\
\hline
$ -z{{E}_{y}}(z,\omega )$ & $E_{\phi ,F}^{inh,n}(r,\omega )$    \\
\hline
$z{{H}_{x}}(z,\omega )$& $H_{\phi ,F}^{inh,n}(r,\omega )$   \\
\hline
$z{{H}_{y}}(z,\omega )$ & $H_{\phi ,A}^{inh,n}(r,\omega )$          \\
\hline
$z{{E}_{x}}(z,\omega )$ & $E_{\phi ,A}^{inh,n}(r,\omega )$  \\
\hline
$\frac{\mu (z,\omega )}{z}$ & $\mu (r,\omega )$     \\
\hline
$\frac{\varepsilon(z,\omega )}{z}$ & $\varepsilon (r,\omega )$ \\
\hline
$\frac{\sigma (z,\omega )}{z}$& $\sigma (r,\omega )$    \\
\hline
$\frac{-1\pm \sqrt{1+4{{k}_{x}}^{2}}}{2}$ & n    \\
\hline
\end{tabular}
\end{table}
The scattered wave from the structure and the transmitted ones in the core can be easily extracted from the mentioned potentials and their relations with EM fields [34]. In addition, the calculation process for the transmitted and reflected waves from the dual IPL is explained exclusively in [35, 36]. Equating the boundary conditions in the inhomogeneous spherical and planar structures, the rest of the equations needed for completing the duality approach are obtained and presented in the Table 2. 
Where $T^{TE}$  and $T^{TM}$  are transmitted coefficients of TE and TM modes, respectively, and the other parameters can be present in the following manner:

\begin{align}
A={{j}^{-n}}\frac{2n+1}{n(n+1)}{\hat{{{J}_{n}}}}({{k}_{0}}{{R}_{o}}) \\
B=\hat{{H}_{n}}^{(2)}({{k}_{0}}{{R}_{o}})\\
C={{j}^{-n}}\frac{2n+1}{n(n+1)}j{\hat{J_{n}^{'}}}({{k}_{0}}{{R}_{o}}) \\
D=j \hat{{H}_{n}^{'}}^{(2)}({{k}_{0}}{{R}_{o}}) \\
{{M}^{TE}}=\left( {{e}^{j{{k}_{z}}{{R}_{o}}}}+{{\Gamma }^{TE}}{{e}^{-j{{k}_{z}}{{R}_{o}}}} \right) \\
{{N}^{TE}}=\left( {{e}^{j{{k}_{z}}{{R}_{o}}}}-{{\Gamma }^{TE}}{{e}^{-j{{k}_{z}}{{R}_{o}}}} \right) \\
{{M}^{TM}}=\left( {{e}^{j{{k}_{z}}{{R}_{o}}}}+{{\Gamma }^{TM}}{{e}^{-j{{k}_{z}}{{R}_{o}}}} \right) \\
{{N}^{TM}}=\left( {{e}^{j{{k}_{z}}{{R}_{o}}}}-{{\Gamma }^{TM}}{{e}^{-j{{k}_{z}}{{R}_{o}}}} \right) \\
{{\theta }_{t}}={{\sin }^{-1}}(\frac{{{k}_{x}}}{{{k}_{d}}}) \\
{{k}_{z1}}={{k}_{d}}\cos ({{\theta }_{t}})
\end{align}
Where ${{\theta }_{t}}$ is the transmission angle. In addition, ${{\varepsilon }_{d}}$ and ${{\mu }_{d}}$ are  EM parameters of dielectric or metamaterial spherical core. ${{\Gamma }^{TE}}$ and ${{\Gamma }^{TM}}$ are the reflection coefficients for TE and TM modes for planar structure, respectively. Details of the derivation of the duality between two structures are explained completely in the appendix B.\begin{table}[!h]
\renewcommand{\arraystretch}{1.2}
\caption{Comprehensive duality between scattering parameters}
\label{table_2}
\centering
\begin{tabular}{|c||c|}
\hline
IPL & ISL\\
\hline
$-\frac{\left( A{{N}^{TM}}\cos ({{\theta }_{i}})+C{{M}^{TM}} \right)}{\left( B{{N}^{TM}}\cos ({{\theta }_{i}})+D{{M}^{TM}} \right)}$ & ${{a}_{n}}$   \\
\hline
$-\frac{\left( A{{N}^{TE}}\cos ({{\theta }_{i}})+C{{M}^{TE}} \right)}{\left( B{{N}^{TE}}\cos ({{\theta }_{i}})+D{{M}^{TE}} \right)}$ & ${{b}_{n}}$  \\
\hline
$\frac{{T}^{TM}(1-\cos {\theta }_{t}){{e}^{j{{k}_{z1}}{{R}_{i}}}}{{k}_{d}}{\hat{R_{i}^{2}}}}{( \hat{{J}_{n}}({{k}_{d}}{{R}_{i}})+j{{\hat{{J}_{n}^{'}}}({{k}_{d}}{{R}_{i}}))}}$ & ${{c}_{n}}$          \\
\hline
$\frac{{{T}^{TE}}( 1-\cos {{\theta }_{t}} ){{e}^{j{{k}_{z1}}{{R}_{i}}}}{{k}_{d}}\hat{R_{i}^{2}}}{(\hat{{J}_{n}}({{k}_{d}}{{R}_{i}})+j{{\hat{{J}_{n}}^{'}}({{k}_{d}}{{R}_{i}}))}}$ & ${{d}_{n}}$  \\
\hline
\end{tabular}
\end{table}
Through previous sections, the problem of EM scattering from a spherical core coated with a radially inhomogeneous layer was addressed. In the following, analysis of scattering from an inhomogeneous sphere is presented as another set of applicable RISSs.
 Using far-field approximation for Riccati-Hankel, the bistatic radar cross section is defined as follows [34]:
\begin{equation}
\begin{split}
{{\sigma }_{(bistatic)}}=\underset{r\to \infty }{\lim }[4\pi {{r}^{2}}\frac{{{\left| {{E}^{s}} \right|}^{2}}}{{{\left| {{E}^{inc}} \right|}^{2}}}  \\ 
 =\frac{{{\lambda }^{2}}}{\pi }[ {{\cos }^{2}}\phi {{|{{A}_{\theta }}|}^{2}}+{{\sin }^{2}}\phi {{|{{A}_{\phi }}|}^{2}}]
\end{split}
\end{equation}
Where:
\begin{align}
{{|{{A}_{\theta }}|}^{2}}={{|\sum\nolimits_{n=1}^{\infty }{{{j}^{n}}}[ {{c}_{n}}\frac{dP_{1}^{n}(\cos \theta )}{d\theta }+{{d}_{n}}\frac{P_{1}^{n}(\cos \theta )}{\sin \theta } ] |}^{2}} \\
{{|{{A}_{\phi }}|}^{2}}={{|\sum\nolimits_{n=1}^{\infty }{{{j}^{n}}}[ {{c}_{n}}\frac{P_{1}^{n}(\cos \theta )}{\sin \theta }+{{d}_{n}}\frac{dP_{1}^{n}(\cos \theta )}{d\theta }]|}^{2}}
\end{align}
\subsection{Conditions of achieving the valuable, accurate, and converged solution}
In this section, the conditions in which the differential system of equations (20)-(23) is solvable leading to valid, accurate, and unique converged solution, is discussed.   
Based on the first order Cauchy-Kowalevski theorem in mathematics [37], the above-mentioned system of equations will give a convergent unique solution when the coefficients at all the points on the arbitrary interval are continuous. Thus, in this article, it is assumed that the EM parameters of inhomogeneous material and consequently the coefficients in the equations (20)-(23) and (24)-(27) are continuous the all points on the interval $z=[{{R}_{i}},{{R}_{o}}]$.
Analyzing EM scattering from inhomogeneous sphere, owing to the presence of r=0 in the solution domain, the term (${}^{1}/{}_{r}$ ) becomes singular. To overcome this drawback, the spherical homogeneous dielectric metamaterial core with near zero radius coated with inhomogeneous layers is considered as an equivalent problem similar to [20]. The validity of the proposed approach for inhomogeneous sphere is verified for the proposed new class of lens antenna.
\section{EXAMPLES AND DISCUSSIONS}
In this section, the validity of the proposed approach for the special type of RISS with analytical and closed-form solution is verified. A spherical PEC core coated with a radially inhomogeneous layer with the following constitutive parameters is considered:
\begin{align}
\varepsilon(r)={{\varepsilon}_{0}}{{K}_{1}}{{e}^{-K_{2}r}} \\
\mu(r)={{\mu}_{0}}{{K}_{3}}{{e}^{-K_{2}r}}
\end{align}
Where ${{K}_{1}}$,${{K}_{2}}$, and ${{K}_{3}}$ are equal to 6, 1, and 1, respectively. The internal and external radii of the coating are $\lambda$ (wave length) and $2\lambda$, respectively. The exact form of the radial and frequency dependency for the auxiliary potentials of the mentioned inhomogeneous medium is presented in the following manner. The proposed general form is derived simply by solving the differential equations (17)-(18) [38].
\begin{equation}
\begin{split}
&F_{r}^{inh,n}(r,\omega )= \\
&\sqrt{r}{{e}^{\frac{{{K}_{2}}r}{2}}}{{c}_{n}}{{J}_{\left( \frac{1}{2}\sqrt{4n(n+1)+1} \right)}} -\frac{1}{2}i\left( {{K}_{2}}^{2}-4{{\omega }^{2}}{{\mu }_{0}}{{\varepsilon }_{0}}{{K}_{1}}{{K}_{3}} \right)r+\\
&\sqrt{r}{{e}^{\frac{{{K}_{2}}r}{2}}}{{d}_{n}}{{Y}_{\left( \frac{1}{2} \sqrt{4n(n+1)+1} \right) }} -\frac{1}{2}i\left( {{K}_{2}}^{2}-4{{\omega }^{2}}{{\mu }_{0}}{{\varepsilon }_{0}}{{K}_{1}}{{K}_{3}} \right)r  
\end{split}
\end{equation}
\begin{equation}
\begin{split}
&A_{r}^{inh,n}(r,\omega )= \\
&\sqrt{r}{{e}^{-\frac{{{K}_{2}}r}{2}}}{{a}_{n}}{{J}_{\left( \frac{1}{2}\sqrt{4n(n+1)+1}\right)}} -\frac{1}{2}i\left( {{K}_{2}}^{2}-4{{\omega }^{2}}{{\mu }_{0}}{{\varepsilon }_{0}}{{K}_{1}}{{K}_{3}} \right)r+\\
&\sqrt{r}{{e}^{-\frac{{{K}_{2}}r}{2}}}{{b}_{n}}{{Y}_{\left( \frac{1}{2} \sqrt{4n(n+1)+1} \right)}} -\frac{1}{2}i\left( {{K}_{2}}^{2}-4{{\omega }^{2}}{{\mu }_{0}}{{\varepsilon }_{0}}{{K}_{1}}{{K}_{3}} \right)r \\ 
\end{split}
\end{equation}
The structure is illuminated by a plane wave with the strength of electric field 1V/m. Fig. 3 a and b show the bistatic radar cross sections in E-plane (x-z) and H-plane (y-z), which are calculated by the exact solution, the proposed approach, and the multilayer homogeneous equivalent approach. In the multilayer equivalent approach, the inhomogeneous layer is modeled by 60 homogeneous layers with the same thickness and the parameters of each layer are calculated using the step-index approximation. In the computations with finite difference (FD) and Taylor methods using the proposed duality approach, the number of subdivisions for FD calculations is M=50. In addition, in the Taylor series method, the number of terms for the truncated Taylor series is N=40.
As it is observed from the figures, one can easily see that there is a good agreement between the obtained results from the presented method and the exact solution. 
To determine the convergence of the proposed method to the exact closed form solution, the error function is defined as follows.
\begin{equation}
\begin{matrix}
   mean  \\
   relative  \\
   error  \\
\end{matrix}=\frac{1}{180}\left( \begin{matrix}
   \sum\limits_{\theta =0}^{180}{\left| \frac{{{\sigma }_{Eplane,approximate}}-{{\sigma }_{Eplane,exact}}}{{{\sigma }_{Eplane,exact}}} \right|}+  \\
   \sum\limits_{\theta =0}^{180}{\left| \frac{{{\sigma }_{Hplane,approximate}}-{{\sigma }_{Hplane,exact}}}{{{\sigma }_{Hplane,exact}}} \right|}  \\
\end{matrix} \right)
\end{equation}
The computation time and error of the employed methods are compared in Table 3. 
\begin{table}[!t]
\renewcommand{\arraystretch}{1.3}
\caption{ Comparison of the computational efficiency of duality approach, exact solution, and multilayer equivalency method in example 1.}
\label{table_3}
\centering
\begin{tabular}{|c||c||c|}
\hline
 ~ & Computation time (s) & Relative error (percent)\\
\hline
Duality-FD & 32 & 1.783  \\
\hline
 Duality-Taylor & 27 & 1.014 \\
\hline
Multilayer & & \\ homogeneous & 132 & 3.826\\ equivalent  & &  \\
\hline
\end{tabular}
\end{table}
Comparing the results simply and clearly confirms that the proposed method is fast and accurate.
\begin{figure}[!t]
\centering
\includegraphics[width=3.2in]{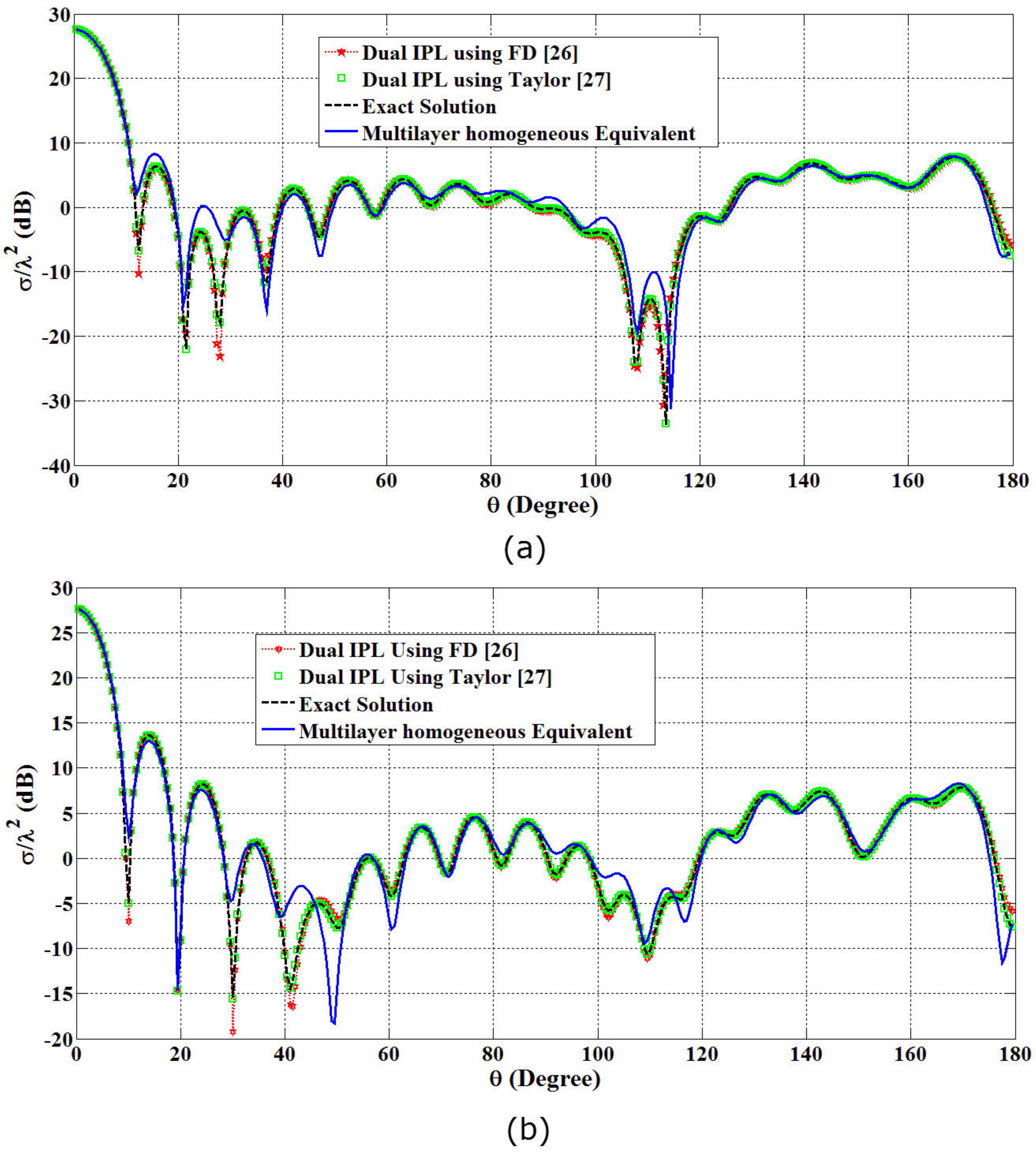}
\DeclareGraphicsExtensions.
\caption{Radar cross section, pertaining to PEC coated with a radially inhomogeneous layer with EM parameters shown in (57) and (58), calculated by the presented method, the exact solution and the multilayer equivalent method in E-plane (a) H-plane (b).}
\label{fig_1}
\end{figure}
\subsection{Designing a new class of lens antenna}
In this section, a new class of lens antenna including inhomogeneous dielectric spheres is proposed for the first time. The proposed lens antenna is based on the physical concept of gradient refractive index material, which is the main principle for all the spherical lens antennas that have been introduced. The refractive index is gradually increased from the value of the surrounding medium permittivity to the maximum value at the center of the inhomogeneous sphere.
The general form of electric permittivity of the proposed lens is presented as follows.
\begin{equation}
{{\varepsilon }_{r}}(r)=\frac{(1+{{A}_{1}})}{(1+{{A}_{1}}{{(\frac{r}{{{R}_{o}}})}^{{{B}_{1}}}})}
\end{equation}
Where ${{A}_{1}}$ and ${{B}_{1}}$ are unknown constant coefficients that are calculated as outputs of optimization algorithms.
Relation (50) simply and clearly shows that the profile of relative permittivity in the inhomogeneous sphere is in complete agreement with the physical concept of the gradient refractive index material in the design of lens antenna.
In this study, the radius of the lens antenna assumed to be 0.6 m at the frequency of 1GHz (or $2 \lambda$).
\begin{figure}[!t]
\centering
\includegraphics[width=3.5in]{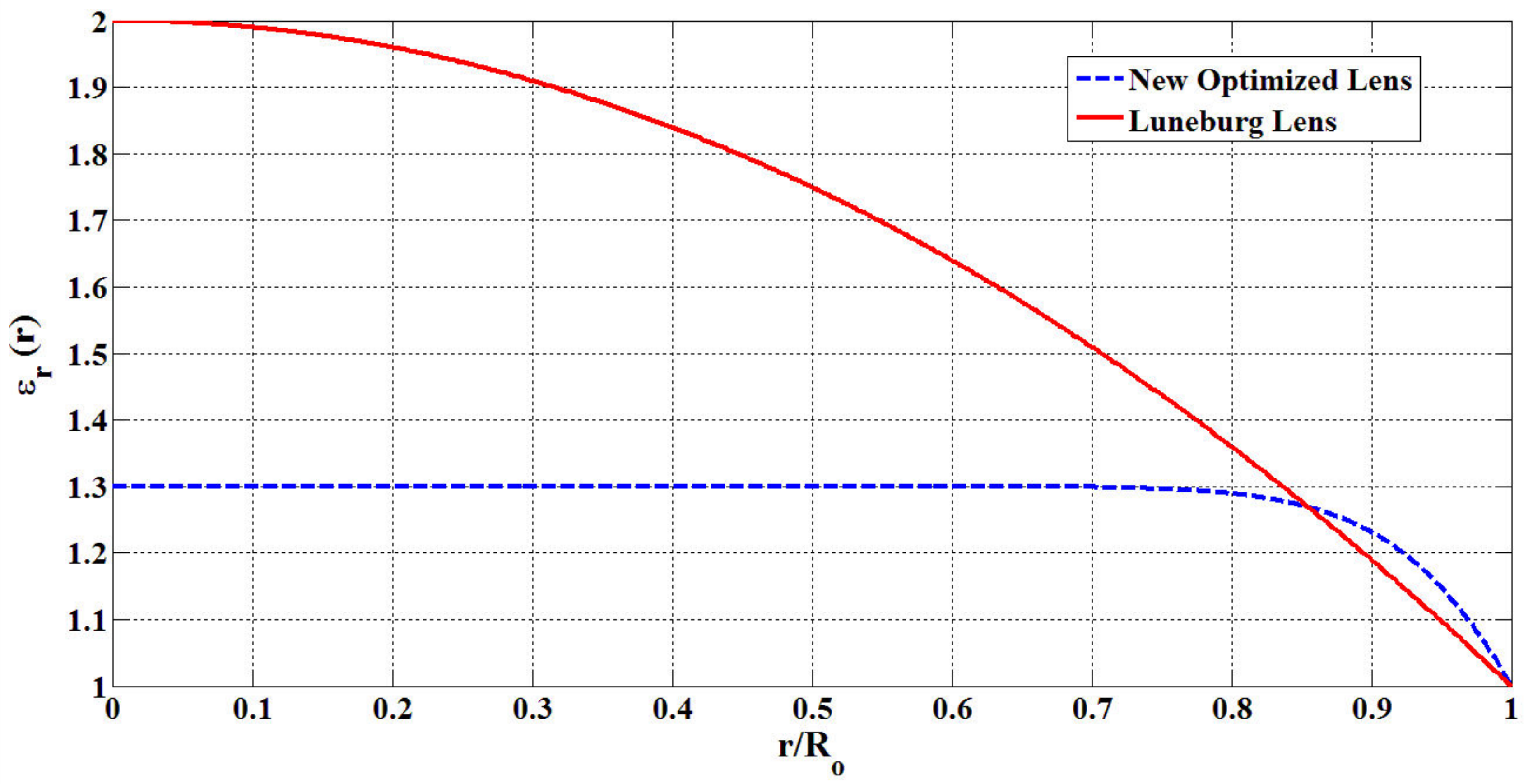}
\DeclareGraphicsExtensions.
\caption{The inhomogeneous profile for relative permittivity obtained by GA corresponding to the optimized lens antenna.}
\label{fig_4}
\end{figure}
As the gradient refractive index concept is proposed independent of the excitation [24-26], a plane wave as a classical source is used. The main goal in the design of a lens antenna is maximum focusing and minimum side-lobe level. Based on the two presented factors, the fitness function for optimization is presented as follows [24]:
\begin{equation}
Fitness({{\varepsilon }_{r}}(r),{{A}_{1}},{{B}_{1}})=\alpha \frac{1}{G(\theta =0)}+\beta G(\theta )\left| _{Sidelobe\,Region} \right.
\end{equation}
Where $\alpha$ and $\beta$ are weight coefficients and is the Gain value at the desired angle.
\begin{figure}[!t]
\centering
\includegraphics[width=3.2in]{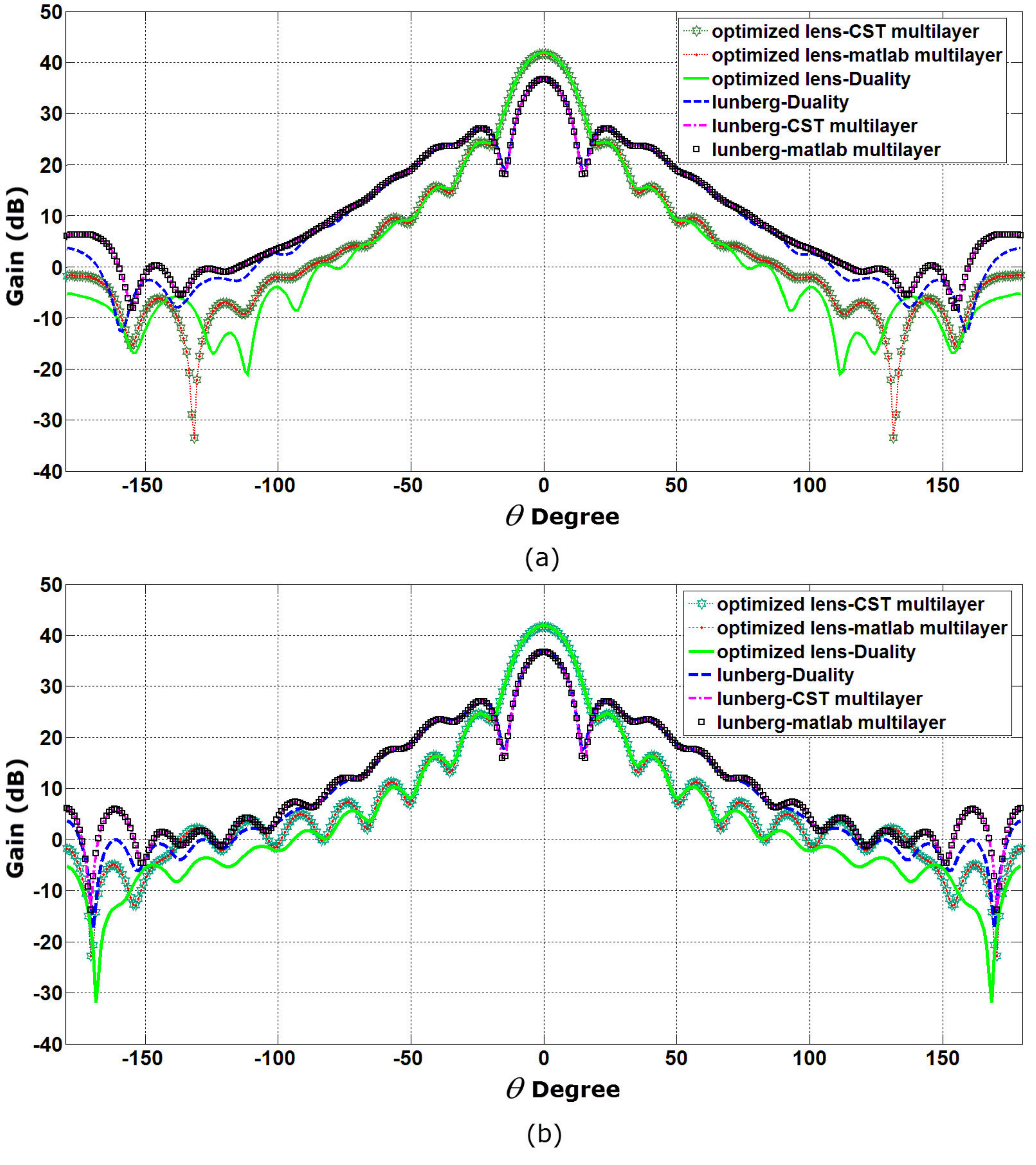}
\DeclareGraphicsExtensions.
\caption{The Gain, pertaining to Luneburg and optimized lens antenna with EM parameters shown in (63), calculated by the presented method, and the multilayer equivalent method in E-plane (a) H-plane (b).}
\label{fig_1}
\end{figure}
Where $\alpha$ and $\beta$ are weight coefficients and $G(\theta )$ is the Gain value at the desired angle.
Genetic algorithm (GA) optimization is used in the proposed method for designing the lens antenna in this section. GA is a favorite global optimization approach that is used commonly. . The $A_1=0.3$ and $B_1=16$ are the result of the GA. The electric permittivity profiles and antenna gain related to these obtained values and applicable Luneburg lens are presented in Fig. 4, and Fig.  5 a, b, respectively. To compute the gain of Luneburg and the proposed lens antennas’, we used the multilayer equivalent method in both the MATLAB and CST softwares along with the proposed method. In this manner, the validity of the proposed method was verified for the cases that do not have an exact solution, despite the presented example in the previous section. It can be clearly seen that the suggested lens has a 5-dB gain enhancement in the main lobe with a 4-dB side lobe level reduction. This design approach can be promising for designing a new class of lenses in a similar manner in future research.
\section{CONCLUSION}
In this contribution, for the first time, an interesting novel duality-based approach is presented for solving the problem of the EM scattering from RISSs. The duality between the wave equations and boundary conditions in the spherical and planar structures was used. The method is exact and without any approximation and is based on strong mathematical and physical background for studying inhomogeneous structures in engineering electromagnetics. The validity of the proposed method was verified through the comprehensive example. Comparing the results obtained from the proposed approach, the exact solution and other commonly used methods in the literature showed that this method is simple, fast, accurate, and valid for all continuous EM profiles. The proposed method substitutes a complex problem in spherical geometry that has a large computational load, with a simple, well-posed problem in planar geometry having less time cost. In addition, the mentioned theory gives us a physical vision of the collision of wave and scattering object geometries.  In the final section, to show the applicability of the method, as a novel application, that is, a special type of dielectric inhomogeneity was introduced to achieve optimum performance for a new class of lens antenna based on the gradient refractive index principle. The proposed method and other conventional approaches are used to confirm the applicability and strong performance of the proposed lens antenna relative to the Luneburg lens antenna.

\appendix

\section{}\label{AppA}

The auxiliary potentials and their relations with electromagnetic fields in a source-free radially inhomogeneous spherical medium ($\nabla .B=0$, $\nabla .D=0$) that can be shown in the following manner.
\begin{align}
D={{\varepsilon }_{comp}}(r,\omega )E=-\nabla \times F \\
B=\mu (r,\omega )H=\nabla \times A \\
{{{\vec{E}_{F}}^{inh}}}=-\frac{1}{{{\varepsilon }_{comp}}(r,\omega )}\nabla \times {{{\vec{F}^{inh}}}} \\
{{\vec{E_{A}}^{inh}}}=-\nabla \psi _{e}^{inh}-j\omega {{{\vec{A}^{inh}}}} \\
{{{\vec{H}_{A}}^{inh}}}=\frac{1}{\mu (r,\omega )}\nabla \times {{{\vec{A}^{inh}}}} \\
{{\vec{H}_{F}^{inh}}}=-\nabla \psi _{m}^{inh}-j\omega {{{\vec{F}^{inh}}}} 
\end{align}
As the inhomogeneity is only toward the radial axis and according to spherical symmetry of the problem, the angular dependency of auxiliary vector and scalar potentials are considered similar to those of the previous investigations based on the presented spherical harmonics in (5)-(7). By placing the introduced angular dependency, the following conclusions can be made for the radial dependency of auxiliary potentials.
\begin{align}
\psi _{e}^{inh}=-\frac{1}{j{{\varepsilon }_{comp}}(r,\omega )}\frac{\partial }{\partial r}(\frac{{{A}^{inh}}}{\mu (r,\omega )}) \\
\psi _{m}^{inh}=-\frac{1}{j\mu (r,\omega )}\frac{\partial }{\partial r}(\frac{{{F}^{inh}}}{{{\varepsilon }_{comp}}(r,\omega )})
\end{align}
According to the above equations, the general form of spatial dependency of EM fields can be derived from equations (10)-(15). Replacing the derived spatial dependency in the Maxwell equations, the radial component is obtained in terms of (with respect to) transverse components as follows.
\begin{align}
E_{r}^{inh}=\cos \phi \sum\nolimits_{n=1}^{\infty }{E_{r}^{inh,n}(r,\omega )P_{1}^{n}(\cos \theta )}\\
H_{r}^{inh}=\sin \phi \sum\nolimits_{n=1}^{\infty }{H_{r}^{inh,n}(r,\omega )P_{1}^{n}(\cos \theta )} \\
H_{r}^{inh,n}(r,\omega )=\frac{n(n+1)}{j\omega r\mu (r,\omega )}E_{\phi ,F}^{inh,n}(r,\omega )	\\
E_{r}^{inh,n}(r,\omega )=\frac{n(n+1)}{j\omega r{{\varepsilon }_{comp}}(r,\omega )}H_{\phi ,A}^{inh,n}(r,\omega )
\end{align}

\section{}
In this section, the procedure of obtaining Table 2 relations is explained. As it was mentioned previously in the manuscript, extraction of Table 2 equations is based on equating boundary conditions of spherical and planar structures. Therefore, boundary conditions for spherical structure with various types of cores is presented in the following manner:\\
Boundary conditions at  $r={{R}_{o}}$:
\begin{align}
E_{\phi ,A}^{inh,n}({{R}_{o}},\omega )=\frac{{{E}_{0}}}{{{k}_{0}}{{R}_{o}}}
\left (\begin{matrix}
 \frac{2n+1}{n(n+1)}j{\hat{J_{n}^{'}}}({{k}_{0}}{{R}_{o}}) \\ 
 +j{a}_{n}{\hat{{H}_{n}^{(2)}}^{'}}({{k}_{0}}{{R}_{o}})
\end{matrix}\right) \\
H_{\phi ,A}^{inh,n}({{R}_{o}},\omega )=\frac{{{E}_{0}}}{{{\eta }_{0}}{{k}_{0}}{{R}_{o}}}\left( \begin{matrix}
 \frac{2n+1}{n(n+1)}{\hat{{{J}_{n}}}}({{k}_{0}}{{R}_{o}}) \\ 
 +{{a}_{n}}{\hat{{H}_{n}^{(2)}}}({{k}_{0}}{{R}_{o}}) \\ 
\end{matrix}\right) \\
E_{\phi ,F}^{inh,n}({{R}_{o}},\omega )=\frac{{{E}_{0}}}{k{{R}_{o}}}\left( \begin{matrix}
 \frac{2n+1}{n(n+1)}{\hat{{{J}_{n}}}}({{k}_{0}}{{R}_{o}}) \\ 
 +{{b}_{n}}{\hat{{H}_{n}^{(2)}}}({{k}_{0}}{{R}_{o}}) \\ 
\end{matrix} \right) \\
H_{\phi,F}^{inh,n}({{R}_{o}},\omega )=\frac{{{E}_{0}}}{{{\eta }_{0}}k{{R}_{o}}}\left( \begin{matrix}
\frac{2n+1}{n(n+1)}j{\hat{J_{n}^{'}}}({{k}_{0}}{{R}_{o}}) \\ 
+j{b}_{n}{\hat{{H}_{n}^{(2)}}^{'}}({{k}_{0}}{{R}_{o}}) \\
\end{matrix}\right)
\end{align}
Boundary conditions at $r={{R}_{i}}$:\\
\textbf{Dielectric or metamaterial core:}
\begin{align}
E_{\phi ,A}^{inh,n}({{R}_{i}},\omega )=\frac{{{E}_{0}}}{{{k}_{d}}{{R}_{i}}}j{{c}_{n}}{\hat{J_{n}^{'}}}({{k}_{d}}{{R}_{i}}) \\
H_{\phi ,A}^{inh,n}({{R}_{i}},\omega )=\frac{{{E}_{0}}}{{{\eta }_{0}}{{k}_{d}}{{R}_{i}}}{{c}_{n}}{\hat{{{J}_{n}}}}({{k}_{d}}{{R}_{i}}) \\
E_{\phi ,F}^{inh,n}({{R}_{i}},\omega )=\frac{{{E}_{0}}}{{{k}_{d}}{{R}_{i}}} {{d}_{n}}{\hat{{{J}_{n}}}}\,({{k}_{d}}{{R}_{i}}) \\
H_{\phi ,F}^{inh,n}({{R}_{i}},\omega )=\frac{{{E}_{0}}}{{{\eta }_{0}}{{k}_{d}}{{R}_{i}}} j{{b}_{n}}{\hat{J_{n}^{'}}}({{k}_{d}}{{R}_{i}})
\end{align}
\textbf{PEC core:}
\begin{align}
E_{\phi ,A}^{inh,n}({{R}_{i}},\omega )=E_{\phi ,F}^{inh,n}({{R}_{i}},\omega )=0 
\end{align}
\textbf{PMC core:}
\begin{align}
H_{\phi ,A}^{inh,n}({{R}_{i}},\omega )=H_{\phi ,F}^{inh,n}({{R}_{i}},\omega )=0 
\end{align}
\textbf{Surface impedance boundary conditions [39]:}
\begin{align}
E_{\phi ,A}^{inh,n}({{R}_{i}},\omega )=-{{\eta }_{s}}H_{\phi ,A}^{inh,n}({{R}_{i}},\omega )=0 \\
E_{\phi ,F}^{inh,n}({{R}_{i}},\omega )=-{{\eta }_{s}}H_{\phi ,F}^{inh,n}({{R}_{i}},\omega )=0 
\end{align}
In addition, boundary conditions for dual planar structure is obtained as follows [35, 36]:\\
Boundary conditions at $z={{R}_{o}}$:\\
for ${TE}^{z}$:
\begin{align}
{{E}_{0}}\left( {{e}^{j{{k}_{z}}{{R}_{o}}}}+{{\Gamma }^{TE}}{{e}^{-j{{k}_{z}}{{R}_{o}}}} \right)={{E}_{y}}({{R}_{o}},\omega )            \\
\frac{{{E}_{0}}}{{{\eta }_{0}}}\cos ({{\theta }_{i}})\left( {{e}^{j{{k}_{z}}{{R}_{o}}}}-{{\Gamma }^{TE}}{{e}^{-j{{k}_{z}}{{R}_{o}}}} \right)={{H}_{x}}({{R}_{o}},\omega ) 
\end{align}
for ${TM}^{z}$
\begin{align}
\frac{{{E}_{0}}}{{{\eta }_{0}}}\left( {{e}^{j{{k}_{z}}{{R}_{o}}}}+{{\Gamma }^{TM}}{{e}^{-j{{k}_{z}}{{R}_{o}}}} \right)={{H}_{y}}({{R}_{o}},\omega )       \\
{{E}_{0}}\cos ({{\theta }_{i}})\left( -{{e}^{j{{k}_{z}}{{R}_{o}}}}+{{\Gamma }^{TM}}{{e}^{-j{{k}_{z}}{{R}_{o}}}} \right)={{E}_{x}}({{R}_{o}},\omega )
\end{align}
Boundary conditions at $z={{R}_{i}}$:\\
\textbf{Dielectric or metamaterial core:}\\
for ${TE}^{z}$:
\begin{align}
{{E}_{0}}{{T}^{TE}}{{e}^{j{{k}_{z1}}{{R}_{i}}}}={{E}_{y}}({{R}_{i}},\omega ) \\
\frac{{{E}_{0}}}{{{\eta }_{0}}}\cos ({{\theta }_{t}})\left( {{T}^{TE}}{{e}^{j{{k}_{z1}}{{R}_{i}}}} \right)={{H}_{x}}({{R}_{i}},\omega )
\end{align}
for ${TM}^{z}$
\begin{align}
\frac{{{E}_{0}}}{{{\eta }_{0}}}{{T }^{TM}}{{e}^{j{{k}_{z1}}{{R}_{i}}}}={{H}_{y}}({{R}_{i}},\omega ) \\ 
{{E}_{0}}\cos ({{\theta }_{t}})\left( {{T }^{TM}}{{e}^{j{{k}_{z1}}{{R}_{i}}}} \right)={{E}_{x}}({{R}_{i}},\omega )
\end{align}
\textbf{PEC core:}
\begin{equation}
{{E}_{y}}({{R}_{i}},\omega )={{E}_{x}}({{R}_{i}},\omega )=0
\end{equation}
\textbf{PMC core:}
\begin{equation}
{{H}_{x}}({{R}_{i}},\omega )={{H}_{y}}({{R}_{i}},\omega )=0 
\end{equation}
\textbf{Surface impedance boundary conditions [39]:}
\begin{align}
{{E}_{y}}({{R}_{i}},\omega )={{\eta }_{s}}{{H}_{x}}({{R}_{i}},\omega ) \\
{{E}_{x}}({{R}_{i}},\omega )=-{{\eta }_{s}}{{H}_{y}}({{R}_{i}},\omega )
\end{align}
Where ${{\eta }_{s}}$is the surface impedance and ${{\varepsilon }_{rd}}$ and ${{\mu }_{rd}}$ are EM parameters of dielectric or metamaterial core.
 By equating boundary conditions of both structures in $z={{R}_{o}}$, $r={{R}_{o}}$,  and $z={{R}_{i}}$, $r={{R}_{i}}$, Table 2 equations and equations (32-41) in the manuscript  was easily obtained. In addition, according to the proposed duality, it can be shown that for the fields at $z={{R}_{i}}$ and $r={{R}_{i}}$, the boundary of both structures are equal for structures with PEC, PMC, and impedance boundary condition core.

\end{document}